\documentclass[aps,twocolumn,superscriptaddress]{revtex4-1}     
\usepackage{graphicx}
\usepackage{amsmath,amssymb}
\usepackage{dsfont}
\usepackage{bm}
\usepackage{color}
\usepackage{MnSymbol, marvosym, wasysym}

\begin{document}

\renewcommand{\vec}[1]{\boldsymbol{#1}}
\newcommand{\up}{{\uparrow}}
\newcommand{\dw}{{\downarrow}}
\newcommand{\pa}{{\partial}}
\newcommand{\pd}{{\phantom{\dagger}}}
\newcommand{\bs}[1]{\boldsymbol{#1}}
\newcommand{\todo}[1]{{\textbf{\color{red}ToDo: #1}}}
\newcommand{\new}[1]{{\color{black}#1}}
\newcommand{\sr}[1]{{\color{black}#1}}
\newcommand{\eps}{{\varepsilon}}
\newcommand{\nn}{\nonumber}
\newcommand{\ie}{{\it i.e.},\ }
\def\eg{\emph{e.g.}\ }
\def\ea{\emph{et al.}}
\def\cf{\emph{c.f.}\ }

\graphicspath{{./}{./figures/}}


\title{Triplet Superconductivity from Nonlocal Coulomb Repulsion\\
in an Atomic Sn Layer Deposited onto a Si(111) Substrate}

\author{Sebastian Wolf}
\affiliation{School of Physics, University of Melbourne, Parkville, VIC 3010, Australia}
\author{Domenico Di Sante}
\affiliation{Department of Physics and Astronomy, University of Bologna, Bologna, Italy}
\affiliation{Center for Computational Quantum Physics, Flatiron Institute, New York, NY, USA}
\author{Tilman Schwemmer}
\affiliation{Institut f\"ur Theoretische Physik und Astrophysik, Universit\"at W\"urzburg,
Am Hubland Campus S\"ud, W\"urzburg 97074, Germany}
\author{Ronny Thomale}	
\affiliation{Institut f\"ur Theoretische Physik und Astrophysik, Universit\"at W\"urzburg,
Am Hubland Campus S\"ud, W\"urzburg 97074, Germany}
\author{Stephan Rachel}
\affiliation{School of Physics, University of Melbourne, Parkville, VIC 3010, Australia}

\date{\today}


\begin{abstract}
Atomic layers deposited on semiconductor substrates introduce a platform for the realization of the extended electronic Hubbard model, where the consideration of electronic repulsion beyond the onsite term is paramount. Recently, the onset of superconductivity at 4.7K has been reported in the hole-doped triangular lattice of tin atoms on a silicon substrate. Through renormalization group methods designed for weak and intermediate coupling, we investigate the nature of the superconducting instability in hole-doped Sn/Si(111). We find that the extended Hubbard nature of interactions is crucial to yield triplet pairing, which is f-wave (p-wave) for moderate (higher) hole doping. In light of persisting challenges to tailor triplet pairing in an electronic material, our finding promises to pave unprecedented ways for engineering unconventional triplet superconductivity.
\end{abstract}

\maketitle



{\it Introduction.---}
Unconventional superconductivity is one of the most active and exciting fields of physics research. The discovery of high-temperature superconductivity in doped cuprate materials in 1986\,\cite{bednorz86zpb189} and in iron-based superconductors 20 years later\,\cite{stewart11rmp1589,mazin-09pc614} mark most notable events and has continuously fueled the search for room temperature superconductors. Similarly, inherent topological superconductivity\,\cite{sato_topological_2017,kallin_chiral_2016,tanaka-12jpsp011013,smidman-17rpp036501} has recently gained high importance, as the quasiparticles at zero energy (referred to as Majorana zero modes) exhibit non-Abelian braiding statistics\,\cite{ivanov01prl268}, rendering them promising candidates for topological qubits and fault-tolerant quantum computing\,\cite{nayak-08rmp1083}. Note that in order to narrow down the meaning of topological superconductors that is henceforth implied in this manuscript, it should be distinguised from proximity-induced topological superconductors, where in general a conventional superconductor induces topological pairing in an adjacent spin-orbit coupled electron liquid\,\cite{lutchyn-10prl077001,oreg-10prl177002,mourik_signatures_2012,nadj-perge-14s602,palacio-morales_atomic-scale_2019}. The list of candidate materials for  inherent topological superconductivity as a potential Majorana platform is rather short. This is because at least at a moderate level of sophistication where one intends to accomplish an odd number of vortex core Majorana zero modes, one is primarily interested in chiral triplet superconductors, which as a predominant initial challenge brings about the task to identify triplet pairing candidate materials in the first place. Examples which have attracted much attention lately are 
CePt$_{3}$Si\,\cite{bauer-04prl027003,yanase-08jpsp124711}, Cu$_x$Bi$_2$Se$_3$\,\cite{hor-10prl057001,sasaki-11prl217001}, FeSe$_{0.45}$Te$_{0.55}$\,\cite{zhang-18s182,rameau-19prb205117,wnag-20s104} and UTe$_2$\,\cite{ran-19s684}.
The most prominent example, however, had been Sr$_2$RuO$_4$\,\cite{mackenzie-03rmp657} -- for decades the prime candidate for $p$-wave superconductivity. Yet, only recently, it was realized that the Knight shift evidence in favor of triplet pairing had to be reconsidered, now favouring singlet pairing\,\cite{pustogow-19n72} which, in order to comply with the other experimental evidence for strontium ruthenate, has to be described by a two-component complex order parameter~\cite{kivelson-20arXiv,suh_stabilizing_2019,ghosh-21np199}.





Two-dimensional atom lattices on semiconductor substrates are a material platform with a rather simple electronic structure. The adsorption of only 1/3 monolayer of group-IV elements such as Pb and Sn forms a $(\sqrt{3}\times\sqrt{3})R30^\circ$ structure, realizing a triangular lattice. Three out of four of the adatoms' valence orbitals are engaged in covalent back bonds with the substrate, leaving the fourth orbital as a half-filled dangling bond. As a consequence for the electronic structure, a single metallic surface band is present within the substrate's band gap. Such a half-filled surface band is subject to significant electron-electron interactions.
For some of these materials, the presence of non-negligble non-local Coulomb interactions was suggested as the driving mechanism for their charge-ordered ground states (\eg in Pb/Ge(111), Sn/Ge(111) or Pb/Si(111)\,\cite{cortes-13prb125113,hansmann-13prl166401,adler-19prl086401,badrtdinov-16prb224418,tresca-18prl196402}). For Sn/Si(111), \new{it was shown in angle-resolved photoemission spectroscopy (ARPES) that the observed band folding was deemed consistent with an antiferromagnetic ordering of the Sn lattice}, in agreement with theoretical modeling of a Mott insulator\,\cite{li-13nc1620}. Soon after, it was also argued that a Slater-type insulator might be the source of the observed magnetism\,\cite{lee-14prb125439}. In Ref.\,\onlinecite{ming-17prl266802}, spectral weight transfer as well as the formation of a quasi-particle peak was demonstrated experimentally, establishing the Mott picture.
Moreover, non-local Coulomb interactions seem to be non-negligible and play an important role in Sn/Si(111)\,\cite{hansmann-13prl166401,adler-19prl086401}.
Most recently, the onset of superconductivity at $4.7 \pm 0.3$K has been observed in strongly boron-doped Sn/Si(111)\,\cite{wu-19arXiv1912.03242}. The hole-doped silicon substrate was grown first, and the tin surface layer was deposited afterwards, keeping the perfect triangular structure of the tin lattice. Doping levels of up to ten percent could be reached. A sharp tunneling dip near zero bias is observed for low temperatures, compatible with a superconducting gap. It is further shown that the gap can be suppressed with increasing magnetic field strength. Detailed analysis of the superconducting gap deviates from conventional BCS behavior, hinting at the unconventional character of the pairing state\,\cite{wu-19arXiv1912.03242}, in agreement with the Mott insulating state of the undoped material\,\cite{li-13nc1620}. The results reproduce many features of the correlated electron physics seen in complex oxides such as high-$T_c$ square lattice cuprate superconductors. In contrast to cuprates, Sn/Si(111) is much simpler, both chemically and electronically, and would thus provide the cleanest platform for studying superconductivity emerging from a doped Mott insulator\,\cite{lee-06rmp17}.

In this Letter, we investigate the competing pairing channels of correlated electrons in Sn/Si(111). By applying the weak-coupling renormalization group (WCRG) method, we are able to find the leading superconducting instabilities in an analytically controlled way in the limit of weak interactions. While correlated electrons on a hexagonal lattice have a generic propensity towards chiral $d$-wave pairing\,\cite{takada-03n53,schaak-03n527,black-schaffer-14jpcm423201,gong-17sae1602579,cao-18prb155145,wu-19prb041117,nandkishore-14prb144501,fischer-14prb020509,disante-19prb201106}, 
we find that the Fermi surface (FS) structure of Sn/Si(111) leads to a competition between singlet and triplet pairing channels. The inclusion of non-local Coulomb interactions strongly suppresses the chiral $d$-wave state, and instead favors $f$-wave and chiral $p$-wave triplet pairing. We further substantiate our findings
by an intermediate coupling analysis through functional renormalization group
(FRG) calculations\,\cite{metzner-12rmp299,platt-13ap453}; \sr{in addition, we can test competing ordering tendencies and rule out other many-body instabilities}. From the synopsis of all results, we find that significant hole-doping in combination with significant non-local Coulomb interactions--as proposed earlier\,\cite{hansmann-13prl166401,adler-19prl086401}--might stabilize topologically non-trivial chiral $p$-wave pairing in Sn/Si(111).


{\it Model and Method.---}
%
%
\begin{figure}[t!]
\centering
\includegraphics[width=0.85\columnwidth]{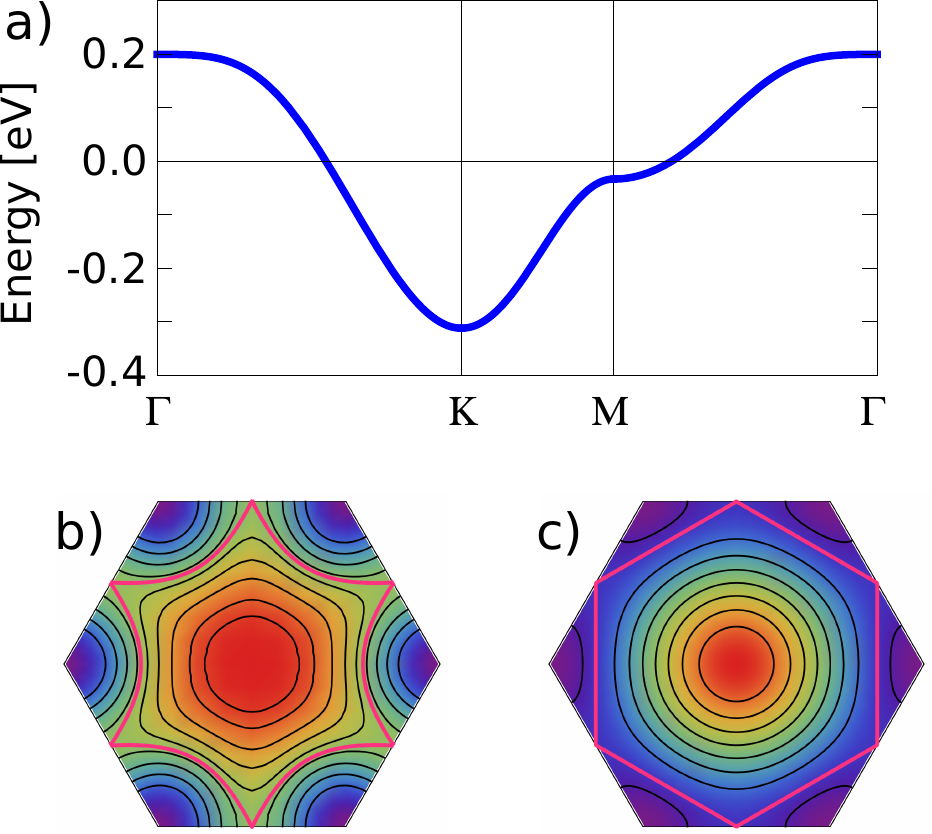}
\caption{(a) Tight-binding bandstructure \eqref{bands} for Sn/Si(111). Fermi surfaces for different fillings are shown for Sn/Si(111) (b) and for the nearest-neighbor triangular lattice (c) as black lines. The van-Hove fillings are shown in red.
}
\label{fig:bands}
\end{figure}
%
We assume that boron-doped Sn/Si(111) is well-described by an extended Hubbard model on the triangular lattice. The electronic bandstructure is derived from {\it Local Density Approximation} (LDA) first-principle calculations\,\cite{li-13nc1620,adler-19prl086401,hansmann-13prl166401}, resulting in a single metallic band well-separated from other bands. A tight-binding fit to the LDA band yields hopping terms up to fourth-nearest neighbors, $\mathcal{H}_0 = \sum_{ij} t_{ij} c_{i\sigma}^\dag c_{j\sigma}^\pd$ and $t_{ij}\equiv t_{|j-i|}$; the bandstructure reads
\begin{equation}\label{bands}
\begin{split}
\eps_{\bs{ k}} =& -2 t_1 \left[ \cos{k_x}+2\cos{\sqrt{3}/2 k_y}\cos{k_x/2}\right] \\
&-2 t_2 \left[ \cos{\sqrt{3}k_y}+2 \cos{2/3 k_x}\cos{\sqrt{3}/2 k_y}\right]  \\
&-2 t_3 \left[ \cos{2 k_x} + 2\cos{k_x}\cos{\sqrt{3}/2 k_y}\right] \\
&- 4 t_4 \left[ \cos{5/2 k_x}\cos{\sqrt{3}/2 k_y} + \cos{2 k_x}\cos{\sqrt{3}k_y} \right. \\
&\qquad\quad \left. + \cos{k_x/2}\cos{3\sqrt{3}/2 k_y}
\right]
\end{split}
\end{equation}
with $t_1 = -52.7$\,meV and $t_2/t_1=-0.389$, $t_3/t_1=0.144$, $t_4/t_1=-0.027$\,\cite{adler-19prl086401} \sr{(see Sec.\,III in the supplement\,\cite{supp} for a discussion about variations of the hopping amplitudes)}. The bandstructure along the high-symmetry path is shown in Fig.\,\ref{fig:bands}\,(a) along with its contour plot in the Brillouin zone in Fig.\,\ref{fig:bands}\,(b) where constant-energy lines correspond to the FSs at different fillings. For comparison, Fig.\,\ref{fig:bands}\,(c) displays the analogous plot for an isotropic, nearest-neighbor triangular lattice.


The extended Hubbard Hamiltonian is given by
\begin{equation}
\label{ham}
\mathcal{H} = \mathcal{H}_0 + U_0 \sum_i n_{i\up} n_{i\dw} + U_1 \sum_{\langle ij \rangle} n_i n_j,
\end{equation}
with $n_{i\sigma}\equiv c_{i\sigma}^\dag c_{i\sigma}^\pd$, $n_i \equiv n_{i\up}+n_{i\dw}$, and the  
local  (nearest neighbour) Hubbard interaction strength $U_0$ ($U_1$). Longer-ranged interactions are generically present, but are often screened and hence mostly assumed subdominant by comparison to $U_0$. 
Notably, the situation is different for 
several group-IV adsorbates on semiconductor substrates many of which possess charge-ordered ground states, which likewise tend to be favoured by long-range Coulomb repulsion~\cite{hansmann-13prl166401,adler-19prl086401}. Sn/Si(111) features a homogeneous charge distribution, but instead orders magnetically, albeit with significant non-local Coulomb interaction being present~\cite{footnote1}: the strongest spectral weight of the single-particle spectral function as measured in ARPES is shifted from the $K$ to and beyond the $M$ point\,\cite{footnote1}, as observed in experiments\,\cite{li-13nc1620} and in agreement with previous theoretical work\,\cite{hansmann-13prl166401}. The experimental data is best described by assuming a ratio of nearest-neighbor and local Hubbard interactions $1/3\leq U_1/U_0\leq 1/2$\,\cite{hansmann-13prl166401,adler-19prl086401}. This appears to be reasonable, and we will consider these parameters for the remainder part of the work\,\cite{footnote3}.

We investigate the superconducting instabilities and form factors of the model Hamiltonian\,\eqref{ham} for the hole-doped case by virtue of the WCRG method\new{\,\cite{raghu_superconductivity_2010,raghu-11prb094518,raghu_effects_2012,wolf_unconventional_2018,Scaffidi17,kiesel-12prb121105,cho-13prb064505}} which builds upon the pioneering work of Kohn and Luttinger\,\cite{kohn-65prl524}.
The main idea is to remain in a regime where a renormalized interaction near the FS can be safely calculated, what can be accomplished by sufficiently small interactions.
%
A standard renormalization group procedure\,\cite{shankar94rmp129} is applied for the remaining effective degrees of freedom, \new{where the weak coupling approximation highlights the renormalization of couplings in the Cooper channel, since superconductivity generically is the leading instability channel in the weak coupling limit.}
In the vicinity of fine tuned points such as van Hove singularities or quadratic band touchings, however, other competing channels can be induced even by infinitesimal interactions, hence this treatment only holds away from these points [which is the case for the studied model \eqref{ham}].
We first calculate the lowest order diagrams shown in Fig.\,S1 \sr{of the supplement}\,\cite{supp}, from which we obtain the effective interaction $U_{\rm eff}$, \new{quantifying the superconducting instabilities. The largest $U_{\rm eff}$ corresponds to the leading instability. $U_{\rm eff}$ is also a measure of the critical temperature.}
%
%
%
%
%
%
%
%
We note that these diagrams contain as integrands the static particle-hole susceptibility $\chi_{\rm ph}$ and the static particle-particle susceptibility $\chi_{\rm pp}$. While the latter diverges at the superconducting instability, the former does not, and is hence evaluated in order to identify the symmetries of the superconducting order parameter. \sr{The methodological steps of the method are explained in detail in the supplement\,\cite{supp}}; \new{we emphasize that the method is asymptotically exact only in the limit of infinitesimal interactions, hence leading to a potentially limited predictive power beyond the weak-coupling regime.} For all WCRG calculations we choose between 228 and 312 patching points on the FS and an adaptive integration grid\,\cite{wolf_unconventional_2018}, with effectively (640)$^2$ points in the Brillouin zone\,\cite{footnote2}.

{\it Results.---}
\begin{figure}[t!]
\centering
\includegraphics[width=0.95\columnwidth]{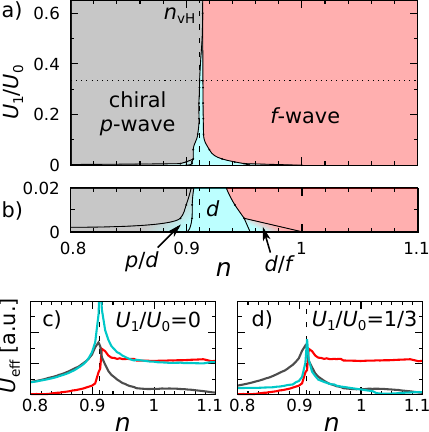}
\caption{(a) Weak coupling phase diagram as a function of non-local Coulomb interaction $U_1/U_0$ vs.\ band filling $n$ for Sn/Si(111). 
For band fillings above the van Hove singularity, triplet superconductivity with $f_{y(3x^2-y^2)}$-wave symmetry  ($B_2$ irrep \new{in red}) is found. For fillings below van Hove (corresponding to hole-dopings larger than 8.5\%) chiral triplet superconductivity with $p_x + i p_y$-wave symmetry ($E_1$ irrep \new{in grey}) is present. (b) Zoom into the phase diagram with very small non-local Coulomb interactions reveal a small area with chiral $d$-wave ($E_2$ irrep \new{in cyan}) pairing as well as regions where $d$-wave is degenerate with $f$-wave (electron-doped) and with $p$-wave (hole-doped). (c) Effective interaction \new{$U_{\rm eff}$} for $U_1=0$: $p$- and $d$-wave pairings ($n<n_{\rm vH}$) and $f$- and $d$-wave pairings ($n>n_{\rm vH}$) are degenerate. (d) For significant $U_1/U_0$ as assumed for Sn/Si(111) $d$-wave pairing is suppressed, leading to chiral $p$-wave and $f$-wave superconductivity. 
}
\label{fig:wcrg-phasedia}
\end{figure}
%
%
%
%
For the particular bandstructure of Sn/Si(111), at $U_1=0$, we find a leading chiral $d$-wave instability only in a comparably small range of dopings $0.9 \leq n \leq 0.95$ [see Fig.\,\ref{fig:wcrg-phasedia}\,(a) and (b)]. 
Instead, for $0.95<n<1$ we find a close-to-degeneracy of $d$-wave and $f$-wave solutions ($E_2$ and $B_2$ irreps). For electron doping $n>1$, the $f$-wave instability is most favorable. Considering hole dopings $n<0.9$,  chiral $d$-wave and chiral $p$-wave solutions are close-to-degenerate ($E_2$ and $E_1$ irreps). The amplitude of the superconducting form factor, $U_{\rm eff}$, is shown in Fig.\,\ref{fig:wcrg-phasedia}\,(c) and reveals the aforementioned competition between the different pairing channels \sr{(the symmetry of the form factor is shown in Fig.\,S2\,\cite{supp})}. This competition between odd- ($f$- and $p$-wave) and even-parity pairing ($d$-wave) is {\it a priori} unexpected as most often chiral $d$-wave is the leading instability for hexagonal tight-binding models\,\cite{black-schaffer-14jpcm423201,cao-18prb155145,wu-19prb041117,nandkishore-14prb144501,fischer-14prb020509,disante-19prb201106}. For comparison, in Fig.\,\ref{fig:bands}\,b) and c) we display the FSs of the Sn/Si(111) band and of an isotropic nearest-neighbor triangular lattice, respectively. Pronounced peaks in the bare susceptibility for the latter are suppressed when switching to Sn/Si(111) where the FS is warped due to longer-ranged hoppings. Thus, even without the inclusion of $U_1$, we can see why singlet-pairing is less favorable than in other hexagonal systems.

Motivated by previous work\,\cite{hansmann-13prl166401,adler-19prl086401}, we assume that non-local Coulomb interactions -- modeled via nearest-neighbor repulsion $U_1$ -- are crucial for boron-doped Sn/Si(111). The presence of $U_1$ further suppresses chiral $d$-wave, and leaves odd-parity $f$-wave pairing ($n>0.91$) or chiral $p$-wave pairing ($n<0.91$) as the leading instability (see Fig.\,\ref{fig:wcrg-phasedia}\,a for $U_1/U_0=1/3$ and Fig.\,S2 in the supplement for $U_1/U_0=1/2$). 
\sr{The suppression of chiral $d$-wave also leads to a significant energetic gap between the leading and the first subleading instability (see Fig.\,\ref{fig:wcrg-phasedia}\,d), indicating the stability of the superconducting ground state with respect to small perturbations as well as finite Coulomb interactions.}
As a central result we identify sufficiently hole-doped Sn/Si(111) as a topological superconductor stabilized by non-local Coulomb repulsion.

\begin{figure}[t!]
\centering
\includegraphics[width=0.99\columnwidth]{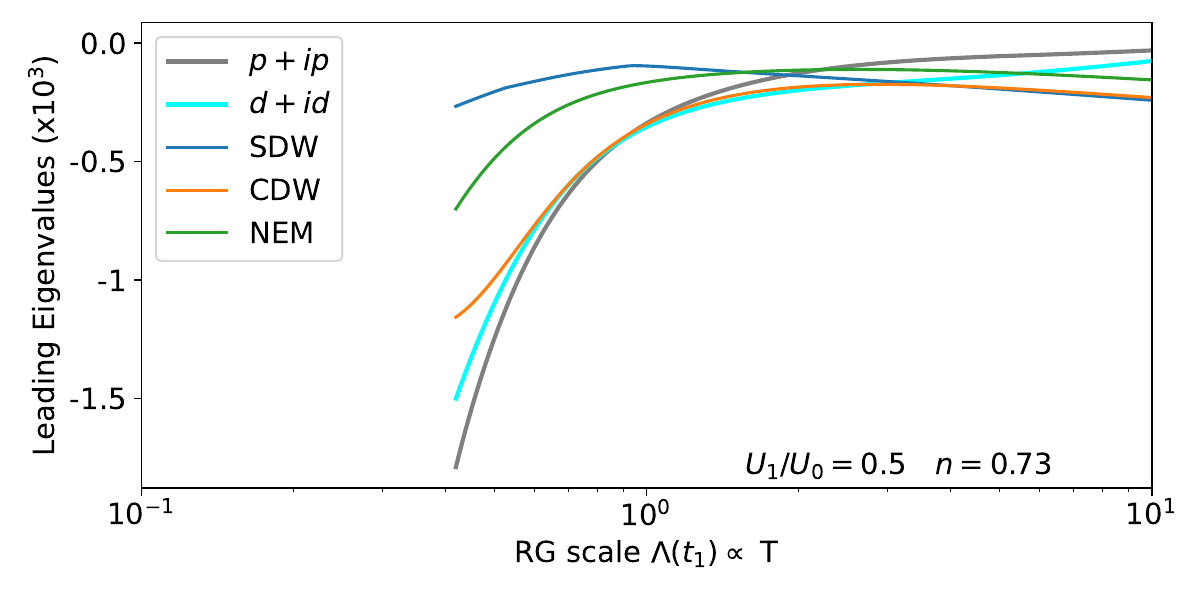} 
\caption{\new{
Leading eigenvalues of the effective two-particle vertex, as a function of $\Lambda$, for superconducting, charge and spin density wave (CDW and SDW, respectively) and nematic (NEM) instabilities. $\Lambda$ is the FRG cutoff energy which is proportional to the temperature. The most negative eigenvalue corresponds to the leading instability.
The chiral triplet $p+ip$ ordering dominates over all other ordering tendencies at low temperatures. Its strongest competitor is chiral $d+id$-wave pairing.
}}
\label{fig:frg}
\end{figure}

\new{We complement our WCRG analysis by numerical FRG calculations based on two key approximations: (i) neglecting the one-particle irreducible (1PI) three particle vertex function and (ii) a Fermi surface based discretization of the 1PI two particle vertex. Since all details of the derivation and limits of the scheme have been extensively reviewed\,\cite{metzner-12rmp299,platt-13ap453,supp} we focus on the key results of our calculation in the following and present a concise summary of our approach in the supplement including Refs.\,\cite{Salmhofer_fRG,dupuis-21pr1}. }
%
As a diagrammatic resummation scheme that includes leading order vertex corrections and treats all two-particle interactions on equal footing, we use the FRG to assess the stability of our WCRG results towards an increase in coupling strength from weak to intermediate coupling. \sr{Moreover, the FRG allows us to rule out other competing many-body instabilities due to its unbiased treatment of the particle-hole and particle-particle channels.} For $U_0=5t_1$ and $U_1=0$, $d$-wave becomes the dominant instability within FRG. This is in line with previous FRG studies: for a Fermiology such as given by Sn/Si(111), sufficient onsite Hubbard coupling strength tends to give preference to singlet pairing. In line with WCRG, however, the addition of $U_1$ indeed can tilt the balance in favor of triplet pairing, as we observe the $d$-wave state in FRG to start to destabilize at finite $U_1$ in favor of $f$-wave pairing at half filling $n\approx 1$ and $p$-wave pairing at $n < 0.75$. Fig.\,\ref{fig:frg} depicts a representative FRG flow diagram for $U_1/U_0=1/2$ where the $p$-wave instability turns out to dominate, and to eventually yield a preferred chiral $p$-wave order parameter in order to maximize condensation energy. 
\sr{FRG further reveals, that competing many-body instabilities such as spin and charge density waves and a nematic phase are subleading compared to superconductivity (Fig.\,\ref{fig:frg}).}
Note that if the interaction scale is increased even further within FRG which would likely be out of the scope of Sn/Si(111), density wave instabilities start to become competitive to the superconducting instability. In total, aside from constraining the preferred doping regime a bit further, WCRG and FRG confirm the propensity of triplet pairing in Sn/Si(111), which in particular turns out to be strong by comparison to previous models studied within FRG.

{\it Discussion.---}
\new{The FS of the bandstructure of Sn/Si(111) is warped due to the longer-ranged hoppings, which leads to a competition between singlet $d$-wave and triplet $p$-wave instabilities. The key for stabilizing triplet $p$-wave superconductivity, however, is the non-local Coulomb interaction $U_1$. When transforming $U_1$ to momentum space, it can be decoupled into contributions associated with the different irreps. They include $p$ ($E_1$ irrep) and $d$-wave ($E_2$ irrep), but not the $f$-wave ($B_2$ irrep) form factors. 
While this explains why the $f$-wave solution in Fig.\,\ref{fig:wcrg-phasedia} c and d is not affected by $U_1$, one would expect $p$ and $d$-wave states to be equally affected by $U_1$ from this analysis.
Representative form factors shown in Fig.\,S2 reveal, however, that the $p$-wave instability has additional nodes. That is the reason why such an ``extended'' $p$-wave is hardly affected by $U_1$, in contrast to the conventional $d$-wave state [see Fig.\,\ref{fig:wcrg-phasedia} c and d]. We stress that extended $p$-wave leads to fully gapped topological super\-con\-ducti\-vity just like the ordinary $p$-wave state.}

Single layers of ordered metal atoms such as Pb and In on Si(111) offer themselves as interesting platforms for conventional electron-phonon mediated superconductivity up to 3K\,\cite{zhang-10np104,noffsinger-11ssc421}.
\sr{We have calculated the electron-phonon coupling for Sn/Si(111), leading to an estimate of $\lambda= 0.07$. This value of $\lambda$ is deeply in the weak coupling regime, and far from the value of $\lambda \sim 1$ typical of high-$T_c$ electron-phonon coupling-mediated superconductors. Details have been delegated to the supplement including Refs.\,\cite{vasp,phonopy,Allen,Pickett}.}
%
%
Moreover, the experimentally observed antiferromagnetic order for the undoped system\,\cite{li-13nc1620} hints at an unconventional pairing mechanism. Our work thus propounds a purely unconventional scenario, where superconductivity emerges upon hole doping of a Mott insulating phase.  

In this work we have neglected the role of spin-orbit coupling (SOC). LDA calculations including SOC show that its effect is rather small. In addition, small to modest contributions of SOC only give rise to some mild mixing of the different pairing channels\,\cite{wolf-20arXiv2004.12624} and the conclusions regarding the dominant superconducting instability will not change.

Our work emphasizes the role of non-local Coulomb interactions in order to stabilize spin-triplet pairing. 
The same had been previously observed for one-band Hubbard models on the square, triangular, and honeycomb lattice\,\cite{wolf_unconventional_2018,wu-19prb041117}.
In a recent paper\,\cite{ghadimi-prb115122} it was found for a Rashba-Hubbard model on the square lattice that the inclusion of $U_1$ leads to an enhancement of triplet pairing. 

Another group-IV adsorbate, which features essentially the same bandstructure as  Sn/Si(111) and is also believed to have significant non-local Coulomb interactions, is Pb/Si(111)\,\cite{adler-19prl086401,cao-18prb155145,hansmann-13prl166401}; it was estimated that $U_1/U_0 = 3/5$ and, according to Fig.\,\ref{fig:wcrg-phasedia}\,(a), we find the same pairing instabilities as for Sn/Si(111). Given the experimental efforts to dope Pb/Si(111), this might be another promising candidate to search for spin-triplet pairing.

\vspace{20pt}

{\it Conclusion.---}
We have shown theoretically that the recently discovered superconductivity in Sn/Si(111) could realize the elusive case of spin-triplet pairing. In contrast to most scenarios on hexagonal lattices where chiral $d$-wave dominates, we show that the bandstructure of Sn/Si(111) leads to a competition between $d$-wave and spin-triplet pairings. We further argue that non-local Coulomb interactions are non-negligible in Sn/Si(111); including them suppresses the chiral $d$-wave state. For moderate hole-dopings we find $f$-wave pairing; most interestingly, hole dopings larger than 8.5\%, as achieved in recent experiments, stabilize the archetypal topologically non-trivial $p+ip$-wave state. Our weak-coupling analysis is backed up by intermediate-coupling FRG simulations which further substantiate our results.
Given the simple chemical and electronic structure of Sn/Si(111) and other adatom lattices on semiconductor substrates, they might provide the cleanest platform for studying (topological) superconductivity emerging from a doped Mott insulator in the future.


\vspace{5pt}
{\it Acknowledgements.---}
We acknowledge discussions with R.\ Claessen, J.\ Sch\"afer, F. Adler, M.\ Laubach, P.\ R.\ Brydon, T.\ L.\ Schmidt and J.\ Makler.
This work is funded by the the Australian Research Council through Grants No.\ FT180100211 and DP200101118, by the Deutsche Forschungsgemeinschaft (DFG, German Research Foundation) through Project-ID 258499086-SFB 1170, the W\"urzburg-Dresden Cluster of Excellence on Complexity and Topology in Quantum Matter – ct.qmat Project-ID 390858490-EXC 2147. The research leading to these results has received funding from the European Union’s Horizon 2020 research and innovation programme under the Marie Sklodowska-Curie Grant Agreement No. 897276. We gratefully acknowledge the HPC facility Spartan hosted at the University of Melbourne and the Gauss Centre for Supercomputing e.V. (www.gauss-centre.eu) for funding this project by providing computing time on the GCS Supercomputer SuperMUC at Leibniz Supercomputing Centre (www.lrz.de). The Flatiron Institute is a division of the Simons Foundation.



\bibliographystyle{prsty} 
\bibliography{snsi111}

\begin{thebibliography}{10}
\bibitem{footnote3}
Previous works assumed values for $U_0$ of the order of $\sim 20
  t_1$\,\cite{hansmann-13prl166401,adler-19prl086401} which is a regime where
  we do not expect superconductivity; thus we consider considerably weaker
  onsite interactions while we assume the ratio $U_1/U_0$ to be unchanged.
\bibitem{footnote2}
We repeated for selected parameter points the calculations with up to 600
  patching points on the FS and for an integration grid with up to (1280)$^2$
  points in the Brillouin zone; the results did not change, proving that the
  chosen resolution is sufficient.
\end{thebibliography}

\vspace{100pt}


\vspace{100pt}

\end{document}


\graphicspath{./figures/}
\newcommand{\todo}[1]{{\color{red}\textbf{#1}}}
\newcommand{\sr}[1]{{\color{black}#1}}
\newcommand{\1}{\hspace*{-1pt}}
\newcommand{\2}{\hspace*{-2pt}}
\newcommand{\3}{\hspace*{-3pt}}
\def\ie{\emph{i.e.},\ }
\def\eg{\emph{e.g.}\ }
\def\ea{\emph{et al.}}
\newcommand{\bs}[1]{\boldsymbol{#1}}
\newcommand{\pa}{\partial}
\newcommand{\up}{\uparrow}
\newcommand{\dw}{\downarrow}
\newcommand{\pd}{\phantom{\dagger}}
\newcommand{\ps}{\phantom{*}}

\renewcommand{\theequation}{S\arabic{equation}}
\renewcommand{\thefigure}{S\arabic{figure}}
\renewcommand{\thetable}{S\arabic{table}}

\title{Supplemental Information:\\Triplet Superconductivity from Nonlocal Coulomb Repulsion\\in an Atomic Sn Layer Deposited onto a Si(111) Substrate}
%

\author{Sebastian Wolf}
\affiliation{School of Physics, University of Melbourne, Parkville, VIC 3010, Australia}
\author{Domenico Di Sante}
\affiliation{Department of Physics and Astronomy, University of Bologna, Bologna, Italy}
\affiliation{Center for Computational Quantum Physics, Flatiron Institute, New York, NY, USA}
\author{Tilman Schwemmer}
\affiliation{Institut f\"ur Theoretische Physik und Astrophysik, Universit\"at W\"urzburg,
Am Hubland Campus S\"ud, W\"urzburg 97074, Germany}
\author{Ronny Thomale}	
\affiliation{Institut f\"ur Theoretische Physik und Astrophysik, Universit\"at W\"urzburg,
Am Hubland Campus S\"ud, W\"urzburg 97074, Germany}
\author{Stephan Rachel}
\affiliation{School of Physics, University of Melbourne, Parkville, VIC 3010, Australia}

\maketitle

%
%
	\section{Weak Coupling Renormalization Group}\label{wcrg}
	Here, only a short overview of the weak coupling renormalization group (WCRG) method will be given, a detailed description can be found in various works  \cite{raghu_superconductivity_2010,raghu_superconductivity_2011,raghu_effects_2012,cho_band_2013,wolf_unconventional_2018}. An important aspect of this method is that it becomes asymptotically exact in the limit of infinitesimal coupling, $U\rightarrow 0$.

	For the WCRG, we consider all terms in the perturbative expansion of the effective two-particle-vertex up to second order in the onsite interaction, $U_{0}$.
	%
	\begin{figure}[b]
	 \centering\includegraphics[width=0.98\columnwidth]{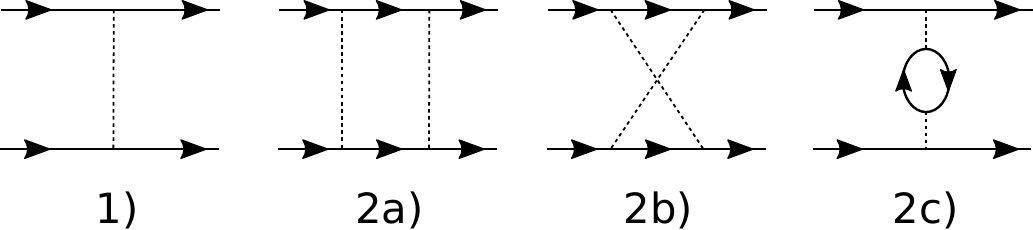}
	 \caption{\label{Fig:App:WCRG_diagram}Non-vanishing spinless particle-hole diagrams of the weak coupling expansion of the effective two-particle-vertex up to second order in the onsite interaction $U_{0}$. Solid lines represent free particle propagators and dashed lines interactions $U_{0}$.}
	\end{figure}
	%
	The renormalization group (RG) flow equation is given by \cite{shankar_renormalization-group_1994,raghu_superconductivity_2010,wolf_unconventional_2018}
	\begin{equation}
	\label{Eq:App:WCRG_flow}
	 \frac{\partial g(k_{2},k_{1})}{\partial \ln(\varepsilon_0/\varepsilon)}=-\int_{\text{FS}}{\text{d}}k_{3}\,g(k_{2},k_{3})g(k_{3},k_{1}),
	\end{equation}
    where $g$ denotes the effective two-particle-vertex, and $k_{1}$ ($k_{2}$) the momenta of the pair electrons with zero total momentum before (after) the scattering process via the interaction $\mathcal{H}-\mathcal{H}_{0}$. The integral on right hand side runs over the Fermi surface (FS). $\varepsilon_0$ denotes the initial infrared cutoff and $\varepsilon$ a lowered cutoff. Note that the final result is still be independent on the initial (arbitrary) cutoff parameter $\varepsilon_0$ \cite{shankar_renormalization-group_1994}.

    In the weak coupling limit we can bypass the procedure of the RG flow, \ie lowering the cutoff $\varepsilon$ in consecutive steps, and obtain the final $g$ directly. All terms which contribute in this manner for the given system are the diagrams shown in Fig.\,\ref{Fig:App:WCRG_diagram}, where we used the spinless variant. Assuming the nearest neighbor interaction to be quadratic in the onsite interaction, \ie
    \begin{equation}
        U_{1}=\alpha \frac{U_{0}^{2}}{W}\ ,
    \end{equation}
    where $W$ denotes the bandwidth, the diagrams 1, 2a, 2b, and 2c are explicitly given by
     \begin{align}
        \label{eq:Gamma_1}
        \frac{g^{(1)}(k_{2},k_{1})}{\tau(k_{1})\tau(k_{2})}&=U_{0}+U_{1}\varepsilon_{1}(k_{2}-k_{1})\nonumber\\
        &=U_{0}+\alpha\frac{U_{0}^{2}}{W}\varepsilon_{1}(k_{2}-k_{1})\ , \\[8pt]
        \label{eq:Gamma_2a}
        \frac{g^{(2a)}(k_{2},k_{1})}{\tau(k_{1})\tau(k_{2})}&=-U_{0}^{2}\int_{\text{FS}}{\text{d}}k\,\frac{1-2f(E(k))}{-2E(k)}\ ,\\[8pt]
        \frac{g^{(2b)}(k_{2},k_{1})}{\tau(k_{1})\tau(k_{2})}&=-U_{0}^{2}\int_{\text{FS}}{\text{d}}k_{3}X_{\text{ph}}(k_{3},k_{1}+k_{2}+k_{3})\ ,     \label{eq:Gamma_2b}\\[8pt]
        \frac{g^{(2e)}(k_{2},k_{1})}{\tau(k_{1})\tau(k_{2})}&=U_{0}^{2}\int_{\text{FS}}{\text{d}}k_{3}X_{\text{ph}}(k_{3},k_{1}-k_{2}+k_{3})\ ,        \label{eq:Gamma_2e}
    \end{align}
    with $\tau(k)$ a $k$-dependent scaling factor,
    \begin{equation}
    \int_{\text{FS}}{\text{d}}k\,\tau^{2}(k)=\int_{\text{BZ}}\frac{{\text{d}}^{2}k}{(2\pi)^{2}}\delta(E)\ ,
    \end{equation}
    the nearest-neighbor kinetic energy $\varepsilon_{1}(k)=\partial\varepsilon(k)/\partial t_{1}$, and the integrand of the particle-hole susceptibility $\chi_{\rm ph}$,
    \begin{equation}
    X_{\text{ph}}(k_{1},k_{2})=\frac{f(E(k_{1}))-f(E(k_{2}))}{E(k_{1})-E(k_{2})}\ .
    \end{equation}
%
%
  $f(E)$ is the the Fermi distribution function. 
  Note that the first order in $U_{0}$ just suppresses the $s$-wave solution.

    The leading superconducting instability is given by the most negative eigenvalue, $\lambda_{\text{min}}$, of $g$. Then, the effective interaction reads $V_{\text{eff}}=\lambda_{\text{min}}/\rho$, where $\rho$ is the total density of states at the Fermi level, and the relation to the critical temperature is given by
    \begin{equation*}
        T_{c}\sim e^{1/(\rho V_{\text{eff}})}=e^{1/\lambda_{\text{min}}}.
    \end{equation*}
    The corresponding eigenfunction, $\psi_{\text{min}}$, yields the form factor of the order parameter, which can be classified by the irreducible representations of the symmetry group of the crystal lattice. 
    \sr{In Fig.\,\ref{fig:op} we show examples of $\psi_{\rm min}$ plotted on the respective FS for the three relevant superconducting instabilities: $p$, $d$ and $f$ wave for the ratios $U_1/U_0=0$ and $1/3$. Note that $p$ and $d$ wave states are two-fold degenerate ($p_x$ and $p_y$, as well as $d_{xy}$ and $d_{x^2-y^2}$); only one of them is shown here.
    \begin{figure}[h!]
\centering
\includegraphics[width=0.99\columnwidth]{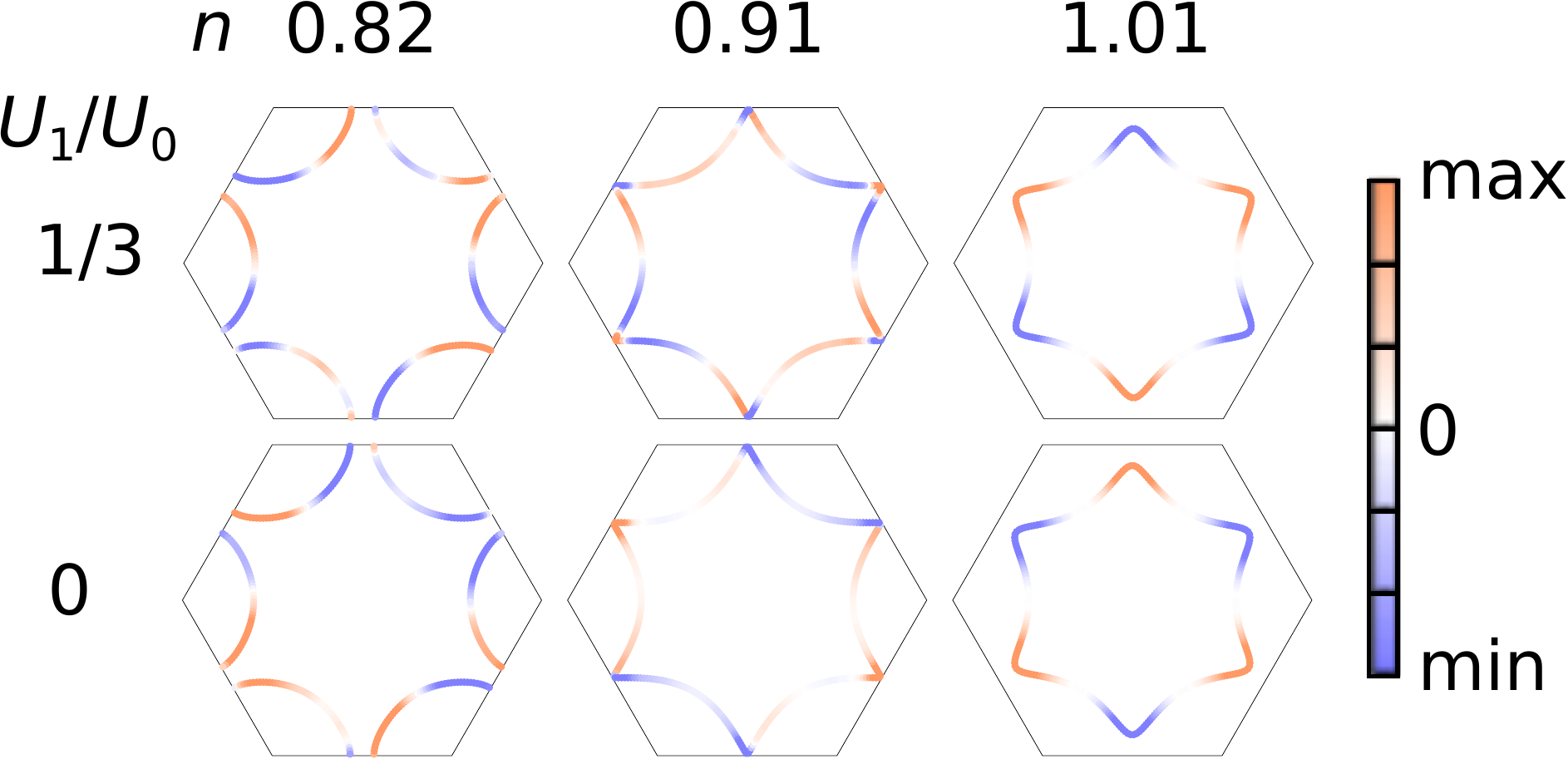}
\caption{\sr{Superconducting order parameter plotted on the corresponding FS. (Left column) One of the two degenerate $p$ wave ($E_1$ irrep) solutions. (Middle column) One of the two degenerate $d$ wave ($E_2$ irrep) solutions. (Right column) $f$ wave ($B_1$ irrep) solution. First (second) row shows results for $U_1/U_0 = 1/3$ ($U_1/U_0=0$).}}
\label{fig:op}
\end{figure}

    }

    Since the three spin orientations of the triplet states are degenerate, it is enough to only calculate $g^{(2b)}$ (yielding the results for spin-singlets and spin-triplets with total spin $S=0$), since the contribution of $g^{(2c)}$ (yielding the results for spin-triplets with total spin $S=\pm1$) is already captured.

    The strength of the second neighbor interaction, $U_{1}$, is adjusted by the parameter $\alpha$, such that
     $U_{1}/U_{0}=\alpha U_{0}/W$. 

\subsection{WCRG phasediagram for $U_1/U_0=1/2$}

\begin{figure}[h!]
\centering
\includegraphics[width=0.7\columnwidth]{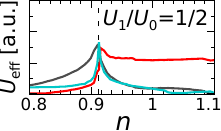}
\caption{Effective coupling for $U_1/U_0=1/2$ (as relevant for SnSi(111)): chiral $d$-wave pairing is suppressed, leading to chiral $p$-wave (below van Hove) and $f$-wave superconductivity (above van Hove). The color code is the same as in Fig.2 in the main paper.}
\label{fig:figs2}
\end{figure}

In the main paper, it is shown that finite $U_1$ leads to a suppression of the chiral $d$-wave singlet pairing state. Instead, triplet pairing (either chiral $p$-wave or $f$-wave) becomes the leading instability. In Fig.\,2d the ratio $U_1/U_0=1/3$ is shown; further increase of $U_1$ does not change the results qualitatively. In Fig.\,\ref{fig:figs2} we present the leading instabilities for $U_1/U_0=1/2$ which are almost the same as in Fig.\,2d.

%
%
    \section{Functional Renormalization Group}
    Similar to section \ref{wcrg}.\ we will only review the basic concepts of the functional renormalization group (FRG) method here, since all technical details have been described in various places in the literature, for example \cite{RevModPhys.84.299,Salmhofer_fRG,platt_functional_2013,dupuis-21pr1} and references therein.
    The starting point for FRG (as for the WCRG) is an itinerant electron model at large energies, as one might obtain it from first-principles calculations. The idea of both approaches is to connect this theory to a low-energy effective model that only takes into account modes close to the Fermi level and absorb the effect of higher energy modes into the renormalized effective interaction.
    In contrast to the WCRG approach, this renormalization group flow is evaluated numerically instead of analytically.
    This is implemented by connecting the effective interaction at a lower energy scale \(\Lambda\) to the effective interaction at a higher energy scale by a set of integro-differential equations, whose analytical form can be calculated exactly.
    However, it turns out that these flow equations do not form a closed set of equations since each equation connects the flow of \(n\)-particle vertex functions with the value of \(n+1\)-particle vertices at that scale.
    Since we are interested in the two-particle vertex function, we make use of the common truncation scheme for this problem, \ie we neglect the effect of three-particle and higher order vertex functions as well as self energy corrections in the flow of the two-particle vertex.
    This truncation was justified in numerous other places \cite{RevModPhys.84.299,platt_functional_2013} and critically shown to capture the relevant interplay between particle-hole (e.g. antiferromagnetic) fluctuations and the particle-particle channel (where antiferromagnetic fluctuations may induce \(d\)-wave superconductivity).
    By using a multi-index notation \(\kappa_i = (k_i,\sigma_i,\omega_i)\)
    The flow equation for the two-particle vertex function can be written as
    \begin{align*}
    \frac{d}{d\Lambda} &\Gamma^{\Lambda}_{\kappa_2,\kappa_3;\kappa_0,\kappa_1} = \sum_{\{\iota_i\}}G^\Lambda_{\iota_0,\iota_2}S^\Lambda_{\iota_1,\iota_3}\Big ( \Gamma^{\Lambda}_{\kappa_2,\kappa_3;\iota_0,\iota_2} \Gamma^{\Lambda}_{\iota_1,\iota_3;\kappa_0,\kappa_1} \\[4pt]
    &- \big [\Gamma^{\Lambda}_{\kappa_2,\iota_3;\kappa_0,\iota_0} \Gamma^{\Lambda}_{\iota_2,\kappa_3;\iota_1,\kappa_1}
    + (\iota_0 \leftrightarrow \iota_1, \iota_2 \leftrightarrow \iota_3)\big]  \\[10pt]
    &+ \big [\Gamma^{\Lambda}_{\kappa_3,\iota_3;\kappa_0,\iota_0} \Gamma^{\Lambda}_{\iota_2,\kappa_2;\iota_1,\kappa_1}
    + (\iota_0 \leftrightarrow \iota_1, \iota_2 \leftrightarrow \iota_3)\big] \Big) \,\text{.}
    \end{align*}
    Note that the three topologically distinct contributions can be further expanded into five diagrams by using the \(SU(2)\) symmetry of the problem, \ie explicitly performing the sum over different spin contributions of the bare (\(G^\Lambda_{\iota_0,\iota_2}=G^\Lambda_{q_0=q_2,\omega_0=\omega_2}\delta_{\sigma_0,\sigma_2}\)) and single-scale propagator \(S^\Lambda_{\iota_1,\iota_3}=S^\Lambda_{q_1,\omega_1}\delta_{\sigma_1,\sigma_3}\).
    The result can be represented by the diagrams shown in Fig.\,\ref{Fig:App:WCRG_diagram}, albeit one needs to convert one of the propagator lines to a single scale propagator in each diagram and take care of the additional graphs produced by this asymmetry.
    Finally we reduce the numerical effort needed to solving this problem by only keeping track of vertex arguments directly on the FS and using a projection scheme to resolve the created ambiguities.
    We implement the flow parameter \(\Lambda\) in the temperature-flow scheme.
    This allows for an unbiased treatment of particle-particle and particle-hole instabilities as well as providing an intuitive picture of understanding the flow as cooling down the system.

    We solve the resulting integro-differential equation using the Euler method.
    The initial condition to the problem is given by enforcing the bare action to equal the renormalized one at an initial cutoff \(\Lambda_{\text{init}}=10\,t_1>U_0\).
    A nice way to analyse the physical implications of the changing two-particle vertex during the flow is to analyse its mean-field components at each point during the flow and plot the effective interaction strength of the leading instabilities as a function of \(\Lambda\).

    Due to the perturbative nature of the approximations made by the truncation of the flow equation, the couplings in the bare vertex will in general grow upon lowering of the cutoff, \ie integrating out high energy modes.
    This eventually results in a breakdown of the truncation conditions and the flow has to be terminated.
    Here this termination condition is fixed to be \(\max \Gamma^{\Lambda}_{\kappa_2,\kappa_3;\kappa_0,\kappa_1} > U_c = 100t_1\).
    Usually the analysis of the mean-field order parameter decomposition of the final vertex identifies a clear instability channel, in the present calculations always in the superconducting channel, and diagonalization analogous to the Cooper vertex analysis for the WCRG method allows us to identify the symmetry of the principal superconducting instability.
    This procedure is performed completely analogous to the one described in Ref.\,\cite{platt_functional_2013} and references therein.
    The critical equation to solve determines not only the irreducible point group representation under which the order parameter \(\Delta(k)\) transforms but also its complete momentum dependence along the FS.

%
%
\sr{\section{Stability of results with respect to changes in the band structure}}

\sr{We have used the tight-binding parameters involving hoppings up to fourth nearest neighbors from Ref.\,\onlinecite{adler-19prl086401}. We note that in Ref.\,\onlinecite{li-13nc1620} independent hoppings are given, which match ours very closely. They contain, however, some small fifth-neighbor hopping $t_5/t_1=0.0318$\,\cite{li-13nc1620}. In contrast, in Ref.\,\onlinecite{hansmann-13prl166401} only hoppings up to third-neighbors are considered, and the hopping ratios $t_2/t_1$ and $t_3/t_1$ are somewhat larger than those found in Refs.\,\cite{adler-19prl086401,li-13nc1620}.

We take these small deviations as a motivation to test the stability of our findings when varying the bandstructure parameters $t_2$, $t_3$ or $t_4$, respectively. Given that the parameter space is three-dimensional, we only vary one parameter at a time. We use the WCRG method to compute the superconducting phase diagram and check the influence of variations of $\pm 10\%$ of the hopping parameters. Specifically, we first vary  $t_2$ while keeping $t_1$, $t_3$, $t_4$ constant and at their original values. Then we repeat this procedure for $t_3$ and $t_4$, respectively.
Such variations in the hopping parameters mostly affect the van Hove singularity and shift it to slightly higher or lower fillings. The superconducting phase diagram changes accordingly. We still find directly at van Hove filling a narrow region with chiral $d$ wave (which further shrinks upon increasing $U_1$), and $f$ wave above and chiral $p$ wave below van Hove filling. With other words, the phase diagram remains invariant with respect to the van Hove singularity. We can further assume, that also combined variations of several hopping parameters would not lead to a qualitatively different outcome (as long as they are small). We have summarized our analysis in Fig.\,\ref{fig:stability}.
\begin{figure}[!t]
\centering
\includegraphics[width=\columnwidth,angle=0,clip=true]{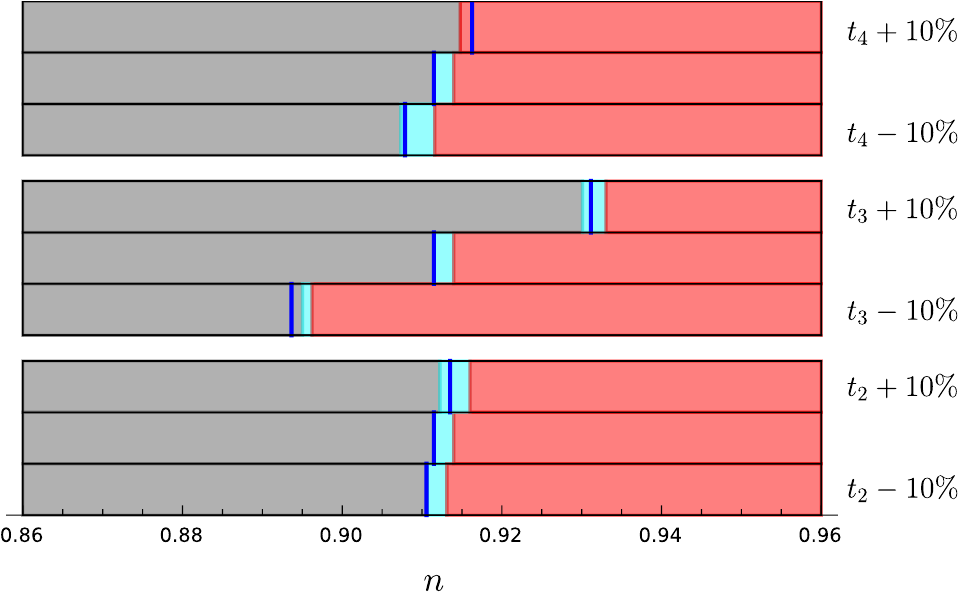}
\caption{\sr{Superconducting phase diagram as a function of filling $n$. Grey (red) correspond to chiral $p$ wave ($f$ wave), while cyan to chiral $d$ wave. The blue vertical lines correspond to van Hove filling. (Top) Variation of fourth-neighbor hopping amplitude $t_4$. The middle line shows the original phase diagram, and the lines above (below) for an increased (decreased) hopping of 10\%. (Center) Variation of third-neighbor hopping amplitude $t_3$. (Bottom) Variation of second-neighbor hopping amplitude $t_2$. For all panels, $U_1/U_0 = 0.1$. Further increase of $U_1$ does not change the overall structure, but shrinks the chiral $d$ wave regions.}
}
\label{fig:stability}
\end{figure}

Interestingly, changes in $t_3$ have the most drastic effect (while variations in $t_2$ and $t_4$ are negligible). If we assume for a moment that the tight binding parameters of Ref.\,\onlinecite{hansmann-13prl166401} were the correct ones, the van Hove singularity would shift to higher fillings $n\approx 0.93$. Thus the chiral $p$ wave solution would also reach higher fillings and require lesser hole-doping to be realized.
}

%
%
%

    \vspace{25pt}
    
    \sr{
    
%
%
    \section{Electron Phonon coupling in $\text{Sn/Si(111)}$}
    
\begin{figure}[!h]
\centering
\includegraphics[width=\columnwidth,angle=0,clip=true]{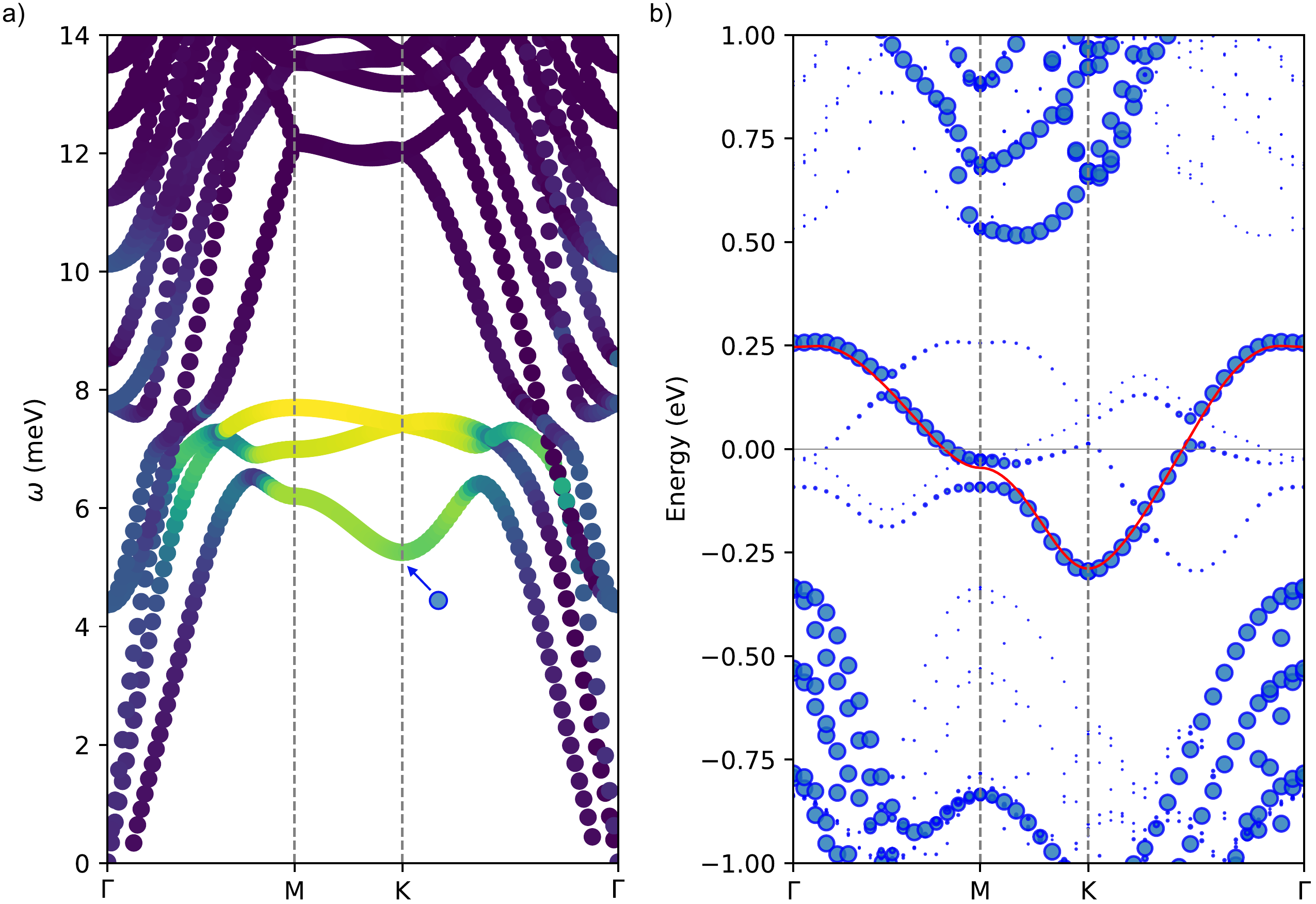}
\caption{\sr{
a) Phonon bandstructure of Sn/Si(111) along the high-symmetry lines, where the yellowish color highlights a strong Sn contribution to the normal modes. b) Electronic bandstructure of Sn/Si(111) when the displacement given by the phonon mode at the bottom of the first phonon branch at the K valley (blue arrow and dot in panel a) is activated. Since this mode requires an enlargement of the unit-cell, the bandstructure is unfolded onto the primitive Brillouin zone in order to compare with the pristine electronic dispersion (red line). The size of the blue dots is a measure of the unfolding weight. The ab-initio and phonon calculations have been performed with the VASP~\cite{vasp} and phonopy~\cite{phonopy} codes, respectively.}
}
\label{ephfig1}
\end{figure}

Here we address the issue of the electron-phonon (EP) coupling in Sn/Si(111). Due to the size of the system, a fully first-principles study is unfeasible. However, if one assumes that the wave vector dependence of the EP coupling is not strong, there is a simple approach using deformation potentials $\mathcal{D} = \Delta E_k/\Delta Q$ due to frozen phonon modes with mode amplitude $Q$. The main idea behind this approach is that a phonon that strongly couples to states nearby the FS produces a large shift $\Delta E_k$ in the electronic dispersion at or near the Fermi level.

The ab-initio phonon dispersion is shown in Fig.~\ref{ephfig1}a) along the Brillouin zone high-symmetry lines. In order to highlight the contribution of the surface Sn atom, the phonon eigenvalues are colored according to the Sn weight in the eigenmodes. One immediately sees that the three Sn phonon bands (there is only one Sn atom per $\sqrt{3}\times\sqrt{3}-$unit-cell) disperse inside a gap into the Si(111) phonon dispersion. Within the deformation potential approximation, the EP coupling of a specific phonon mode is calculated as~\cite{Allen}

\begin{eqnarray}
\label{ephEq1}
\lambda = 2 N(E_F)\left[ \frac{\hbar}{2M_\text{Sn}\omega^2} \right]|\mathcal{D}|^2
\end{eqnarray}

\noindent where $N(E_F)$ is the density-of-states per spin and unit-cell, $M_\text{Sn}$ is the mass of the Sn atom, and $\mathcal{D}$ is the aforementioned deformation potential. Inspecting Eq.~\ref{ephEq1} one also notices that the lowest frequency modes contribute most at the EP coupling.

For this reason, we focus our attention to the lowest branch at the K valley with phonon frequency $\sim 5$ meV, since it is further characterized by a strong out-of-plane displacement of the Sn atom. This is quite relevant in light of the $p_z$ character of the FS, which is indeed expected to react to vertical displacements. We create a supercell with a total of 360 atoms, and the four surface Sn atoms displaced according to the first phonon eigenmode at K. The amplitude $Q$ of the displacement is set for convenience to 20 times the eigenmode (in the linear approximation, the deformation potential is independent of the amplitude), resulting in $Q\sim0.13$~\AA. To infer the shift of the electronic states at the Fermi level, we show in Fig.~\ref{ephfig1}b) the Sn/Si(111) bandstructure unfolded onto the Brillouin zone of the primitive cell (the one containing just one surface Sn atom). By comparing it with the red unperturbed dispersion, we see that the effect of the phonon eigenmode is quite small at the FS, being smaller than $\Delta E_k \sim 20$ meV. It is important to notice here that in typical EP coupling mediated superconductors such as MgB$_2$, $\Delta E_k \sim 1.5$ eV is almost two orders of magnitude larger~\cite{Pickett}. For the specific phonon mode we are considering, we then estimate a deformation potential $\mathcal{D} \sim 0.15$ eV/\AA~($\mathcal{D} \sim 13$ eV/\AA ~for the $E_{2g}$ phonon modes of MgB$_2$). We further compute $N(E_F) \sim 2.32$ eV$^{-1}$ and $M_\text{Sn}\omega^2 \sim 0.78$ eV/\AA$^2$~that plugged into Eq.~\ref{ephEq1} give an estimate of $\lambda \sim 0.07$. This value of $\lambda$ is deeply in the weak coupling regime, and far from the value of $\lambda \sim 1$ typical of high-$T_c$ EP coupling-mediated superconductors. 

}

 \bibliography{bib_incl_frg}